\documentclass[letterpaper, preprint, paper,11pt]{AAS}	

\usepackage{bm}
\usepackage{amsmath}
\usepackage{subfigure}
\usepackage[colorlinks=true, pdfstartview=FitV, linkcolor=black, citecolor= black, urlcolor= black]{hyperref}
\usepackage{overcite}
\usepackage{footnpag}		
\usepackage{mathtools}
\usepackage{amsfonts}
\usepackage{cancel}
\usepackage[numbers,sort]{natbib}
\usepackage{tikz}
\usepackage{listings}
\usepackage{xcolor}

\usepackage{algorithm}
\usepackage{algpseudocode}
\usepackage{changepage} 
\newcommand{\norm}[1]{\left\lVert#1\right\rVert}

\PaperNumber{25-916}

\begin{document}

\title{INCORPORATING THE NONLINEARITY INDEX INTO ADAPTIVE-MESH SEQUENTIAL CONVEX OPTIMIZATION FOR MINIMUM-FUEL LOW-THRUST TRAJECTORY DESIGN}

\author{Saeid Tafazzol\thanks{Graduate Student, Department of Aerospace Engineering, Auburn University.}, and
Ehsan Taheri\thanks{Associate Professor, Department of Aerospace Engineering, Auburn University. AAS Member.}}

\maketitle{}

\begin{abstract}
Successive convex programming (SCP) is a powerful class of direct optimization methods, known for its polynomial
complexity and computational
efficiency, making it particularly suitable for autonomous applications. Direct methods are also referred to as ``discretize-then-optimize'' with discretization being a fundamental solution step. A key step in all practical direct methods is mesh refinement, which aims to refine the solution resolution by enhancing the precision and quality of discretization techniques through strategic distribution and placement of mesh/grid points. We propose a novel method to enhance adaptive mesh refinement stability by integrating it with a nonlinearity-index-based trust-region strategy within the SCP framework for spacecraft trajectory design. The effectiveness of the proposed method is demonstrated through solving minimum-fuel, low-thrust missions, including a benchmark Earth-to-Asteroid rendezvous and an Earth-Moon L2 Halo-to-Halo transfer using the Circular Restricted Three-Body (CR3BP) model.
\end{abstract}

\section{Introduction}

Trajectory optimization plays a critical role in modern space exploration missions, impacting both operational effectiveness and mission feasibility \cite{whiffen2006mystic,trelat2012optimal}. As spacecraft mission profiles continue to grow in complexity \cite{arya2021electric}, developing robust trajectory optimization methods has become increasingly vital for successful orbital operations and cost-effective mission planning \cite{ellison2018application}. Recent advancements in spacecraft propulsion technology, particularly electric propulsion systems, have notably enhanced fuel economy compared to conventional chemical thrusters. The specific impulse of electric propulsion systems surpass traditional chemical propulsion by approximately one order of magnitude, substantially improving spacecraft operational efficiency and mission design flexibility \cite{lev2019technological}. 

On the other hand, the use of electric propulsion introduces unique complexities into the trajectory design task, primarily due to the low thrust levels that necessitate long-duration trajectories encompassing multiple orbital revolutions around a planetary body \cite{petukhov2019application,meng2019low}. Indeed, for very low-thrust planet-centric phases of flights,  trajectories often involve numerous revolutions, sometimes numbering in the hundreds, before reaching their destination \cite{aziz2018low,taheri2021optimization}. Furthermore, depending on the selected optimization objective—such as minimizing propellant usage or reducing flight time—the thruster operation may not be continuous, and the optimal solution frequently comprises alternating thrusting and coasting segments \cite{taheri2018generic}. Determining this alternating sequence, commonly known as the optimal control structure, poses a significant challenge, as the quantity and duration of thrusting and coasting periods remain unknown and must be computed during the optimization process \cite{taheri2023l2}.

Numerical methods for solving spacecraft trajectory optimization, and in general, optimal control problems fall into two main categories: \textit{indirect} and \textit{direct} methods \cite{betts1998survey,trelat2012optimal}. Indirect methods involve deriving and solving boundary-value problems based on necessary optimality conditions from Pontryagin's minimum principle \cite{mall2020uniform,kovryzhenko2023vectorized}. Conversely, direct methods approach trajectory optimization by transforming the original infinite-dimensional optimal control problem into a finite-dimensional nonlinear programming (NLP) problem \cite{sowell2024eclipse}. This transformation, known as transcription, discretizes both the state and control variables across the trajectory, facilitating numerical solutions.

In recent years, however, significant advancements have been made in direct optimization techniques due to the increasing availability of efficient computational algorithms and robust solvers \cite{conway2010spacecraft}. Among direct optimization strategies, convex programming has emerged as a particularly attractive subclass due to the advent of highly efficient convex solvers. These convex formulations enable rapid convergence and reliable solutions, often within seconds or few hundreds of milliseconds (depending on the nature of the problem), making them suitable for real-time guidance and autonomous spacecraft operations \cite{boyd2004convex,wang2024survey}. The key concept here is successive convex programming (SCVX), which iteratively approximates the non-convex NLP through sequential convexification steps. At each iteration, nonlinear constraints and dynamics are linearized around a solution estimate obtained from the previous iteration. Each resulting convex subproblem is solved efficiently, progressively refining the solution and guaranteeing convergence under suitable conditions \cite{wang2018minimum}. For instance, Nurre and Taheri \cite{nurre2022comparison} conducted, for the first time, a comparison between indirect and convex optimization based fuel-optimal low-thrust trajectories. They also solved the optimization problems using Cartesian, spherical and the set of modified equinoctial elements presenting a first comparison of the choice of coordinates and its impact for solving convex optimization problems.

Transcription is a fundamental step within direct methods that enables the translation of the infinite-dimensional optimal control problem into a tractable finite-dimensional NLP problem. Here, the continuous trajectory and control inputs are temporally-discretized into a finite set of discrete points or nodes. While it is a common practice to distribute these discretization nodes uniformly during the maneuver time, this approach can become suboptimal, especially in problems where the optimal control structure exhibits multiple switches (e.g., bang-off-bang thrust arcs) \cite{betts1998mesh}. If the temporal distribution of the mesh/grid points does not align accurately with the time instants associated with control switches, the computed solution may deviate significantly from true optimality.  Because the number of control switches and their time instants are initially unknown, adaptive-mesh refinement methods can be used to dynamically adjust the node placement \cite{koeppen2019fast}, improving accuracy and ensuring optimality \cite{kumagai2024adaptive}. This adaptation is typically implemented through time-dilation variables (adjustable parameters within the convex optimization framework) that  reposition mesh/grid nodes closer to regions exhibiting control switches or significant nonlinear dynamics.

To further enhance the robustness and stability of adaptive mesh refinement, this study integrates the mesh refinement with a trust-region strategy guided by a nonlinearity index \cite{junkins2004nonlinear}, inspired by recent developments outlined in \cite{bernardini2024state}.
Given that linear approximations used in SCVX are inherently local, trust-region constraints are employed to bound the solution update between iterations. However, uniformly sized trust regions may not adequately represent highly nonlinear segments of the trajectory. By quantifying nonlinearity through a nonlinearity index, this method adaptively tightens the trust region specifically in areas of greater nonlinear behavior. This nonlinearity-aware thrust-region method ensures that solutions remain valid approximations of the original nonlinear problem, thus enhancing convergence and solution quality for complex low-thrust spacecraft trajectories. 

The rest of the paper proceeds as follows. We begin by reviewing the fundamentals of successive convex optimization. We then formulate the minimum-fuel trajectory optimization problem and describe our adaptive mesh-refinement approach. Next, we introduce the nonlinearity-index–guided trust-region enhancement. After that, we present numerical results on two benchmark trajectories. Finally, we give conclusions and outline directions for future work.







\section{Background on Successive Convex Optimization}

Leveraging the capabilities of convex optimization, one can adopt robust successive convexification frameworks, as demonstrated in~\cite{szmuk2018successive,mao2018successive}, to address optimal control problems, including those that are inherently non-convex. To achieve this, nonlinear dynamics and constraints are linearized around a reference trajectory. Consider the following nonlinear dynamics:
\begin{equation}
    \dot{\bm{x}} = \bm{f}(\bm{x}, \bm{u}),
\end{equation}
where $\bm{x}$ and $\bm{u}$ are state and control vectors, respectively. The nonlinear dynamics can be linearized around reference values $(\hat{\bm{x}}, \hat{\bm{u}})$ using a first-order Taylor series expansion:
\begin{subequations}
\begin{align}
&\dot{\bm{x}}(t) = \bm{f}(\bm{x}(t), \bm{u}(t)) \approx A_L(t) \bm{x}(t) + B_L(t) \bm{u}(t) + \bm{e}(t), \\
&A_L(t) \coloneq \left. \dfrac{\partial \bm{f}(\bm{x}, \bm{u})}{\partial \bm{x}} \right|_{\hat{\bm{x}}(t), \hat{\bm{u}}(t)},~B_L(t) \coloneq \left. \dfrac{\partial \bm{f}(\bm{x}, \bm{u})}{\partial \bm{u}} \right|_{\hat{\bm{x}}(t), \hat{\bm{u}}(t)},\\
&\bm{e}(t) \coloneq \bm{f}(\hat{\bm{x}}(t), \hat{\bm{u}}(t)) - A_L(t) \hat{\bm{x}}(t) - B_L(t) \hat{\bm{u}}(t).
\end{align}
\label{eq:linear_approx}
\end{subequations}
The next step is to discretize the linearized dynamics and enforce the dynamics as constraints in the optimization problem:
\begin{equation} \label{eq:xpfromxm}
    \bm{x}^+ \approx \bm{x}^- + \int_{t^-}^{t^+} \left[ A_L(t) \bm{x}(t) + B_L(t) \bm{u}(t) + \bm{e}(t) \right] dt
\end{equation}

To express Eq.~\eqref{eq:xpfromxm} as a linear constraint, one can use the state transition matrix (STM) $\Phi_A(t^+, t^-)$, which captures the zero-input solution from $\bm{x}^-$ to $\bm{x}^+$. The STM can be computed via the following differential equation:
\begin{equation}
    \dot{\Phi}_A(t, t^-) = A_L(t) \Phi_A(t, t^-), \quad \Phi_A(t^-, t^-) = I.
\label{eq:stm_ode_cvx}
\end{equation}

Then, the discrete dynamic constraint can be enforced as:
\begin{subequations}
\begin{align}
&\bm{x}^+ = \bar{A} \bm{x}^- + \bar{B} \bm{u}^- + \bar{C}_k \bm{u}^+ + \bar{\bm{e}}, \\
&\bar{A} \coloneq \Phi_A(t^+, t^-),~\bar{B} \coloneq \int_{t^-}^{t^+} \Phi_A(t^+, t) B_L(t) \alpha_k(t) dt, \label{eq:conv_action}\\
&\bar{C} \coloneq \int_{t^-}^{t^+} \Phi_A(t^+, t) B_L(t) \beta_k(t) dt,
\end{align}
\label{eq:discrete_dynamics}
\end{subequations}
In this discretization, a first-order hold is used for control, guided by the interpolation functions $\alpha_k(t)$ and $\beta_k(t)$, where $\alpha(t^-) = \beta(t^+) = 1$ and $\alpha(t^+) = \beta(t^-) = 0$. These are linear functions responsible for interpolating control inputs between discrete time steps.

Observing that the linearization is only valid within a neighborhood of the reference trajectory, a trust region is introduced to ensure that the solution remains bounded and consistent with the original problem. Furthermore, since the linearization serves merely as an approximation of the true nonlinear dynamics, a successive convexification approach is employed. This iterative method refines the reference trajectory until it converges to the solution within the local neighborhood, thereby yielding a local solution to the original non-convex problem. The details of this procedure will be presented in the following section.

\section{Minimum-Fuel Trajectory Optimization Problem and Mesh Refinement}
Let $\bm{x}(t) \in \mathbb{R}^6$ denotes the state of the spacecraft and let $\bm{T}(t) \in \mathbb{R}^3$ denote the control input, which is the total thrust vector. The minimum-fuel trajectory optimization problem becomes,
\begin{subequations}\label{eq:ocp_convex}
\begin{align}
\underset{\bm{T}}{\text{Minimize}}\quad 
J &= \int_{t_0}^{t_f} \|\bm{T}(t)\|_2\,dt,
\label{eq:ocp_convex_a}\\
\text{s.t.,}\quad
\dot{\bm{x}}(t) 
&= \mathbb{A}(\bm{x}) + \mathbb{B}(\bm{x})\frac{\bm{T}(t)}{m(t)},
\quad
\dot{m}(t) = -\frac{\|\bm{T}(t)\|_2}{c},
\label{eq:ocp_convex_b}\\
\|\bm{T}(t)\|_2 &\le T_{\max},
\label{eq:ocp_convex_c}\\
\bm{x}(t_0) &= \bm{x}_0,\quad \bm{x}(t_f) = \bm{x}_f,
\label{eq:ocp_convex_d}\\
m(t_0) &= m_0,
\label{eq:ocp_convex_e}
\end{align} 
\end{subequations}
where $m(t) \in \mathbb{R}^+$ represents its mass. In a fixed-time maneuver, our goal is to guide the spacecraft from the initial state, $\bm{x}_0$, with initial mass, $m_0$, at time $t_0$ to the final state, $\bm{x}_f$, at time $t_f$. The constant $c$ denotes the effective exhaust velocity (i.e., $c = I_\text{sp} g_0$ with $g_0$ denoting the sea-level gravity constant and $I_\text{sp}$ denoting the specific impulse). The algebraic expressions for matrices $\mathbb{A}$ and $\mathbb{B}$ depend on the choice of coordinates (e.g., Cartesian coordinates) and/or element sets (e.g., the set of modified equinoctial elements) \cite{taheri2016enhanced,junkins2019exploration}.

\subsection{Convexification and Discretization} \label{section:convexification}
We adopt an adaptive mesh-refinement strategy, wherein the trajectory is not uniformly divided into equally sized segments. Instead, the optimizer is allowed to adjust the segment spacing to further improve the objective value. This is achieved conveniently by introducing a normalized time span accompanied with time-dilation, which is a common step in solving optimal control problems \cite{sowell2024eclipse}. Following \cite{kumagai2024adaptive}, normalized time $\tau \in [0,1]$ with equally spaced $N$ positions such that $\tau_k = (k/N)$ for $\mathbb{Z}_0^N$ \footnote{$\mathbb{Z}_0^N$ denotes the set of integers from $0$ to $N$, i.e., $\mathbb{Z}_0^N = \{0, 1, \dots, N\}$.}. Furthermore, $s : \mathbb{R} \rightarrow \mathbb{R}$ is defined such that $s(\tau) = dt/ d\tau$, which is called time-dilation variable in \cite{szmuk2020successive}. The equations of motion can be reformulated as,
\begin{align}
    &\bm{x}' \coloneq \dfrac{d\bm{x}}{d\tau} = \dfrac{d\bm{x}}{dt} \dfrac{dt}{d\tau} = \dot{\bm{x}}s = s(\tau)[\mathbb{A}(\bm{x}) + \mathbb{B}(\bm{x}) \dfrac{\bm{T}}{m}],\\
    &m' \coloneq \dfrac{dm}{d\tau} = -\dfrac{s\norm{\bm{T}}_2}{c}. \label{eq:m_prime}
\end{align}

The objective can be re-written as, 
\begin{equation} \label{eq: time_di_obj}
\begin{aligned} 
    J &=  \int_{t_0}^{t_f} \norm{\bm{T}(t)}_2 \, dt = \int_{0}^{1} \norm{\bm{T}(\tau)}_2 \underbrace{\dfrac{dt}{d\tau}}_{s(\tau)} \, d\tau. 
\end{aligned}
\end{equation}

The time-dilation variable, $s$, is introduced so that by modifying it, mesh refinement is enabled for the optimizer. However, given the bilinear term in 
objective \eqref{eq: time_di_obj}, convexification would require linearization in its current form. In \cite{acikmese2007convex, wang2018minimum}, a mathematical trick is used by defining $\tilde{\bm{T}} \coloneqq {\bm{T}}/{m}$ as the acceleration vector and using $z \coloneqq \ln m$ as a substitute to mass. This would solve the problem of having thrust vector and mass, which are both variables to convex optimization, being coupled. In this work, we propose a slightly different version of this mathematical trick to overcome the bilinearity (of the decision variables) appearing in the problem objective. In this paper, propose and define $\tilde{\bm{T}} \coloneqq \bm{T}s/m$. We use $J = \int_{0}^{1} \norm{\tilde{\bm{T}}(\tau)}_2 d\tau$ as the new objective. In \cite{wang2018minimum}, it is shown that the objective with $\tilde{\bm{T}} \coloneqq \dfrac{\bm{T}}{m}$ reparameterization, is equivalent to maximizing the final mass. We follow the same step to prove the same statement for the slightly different problem at hand. Using \eqref{eq:m_prime}:
\begin{equation}
    \underbrace{\int_{t_0}^{t_f} \dfrac{\dot{m}(t)}{m(t)}dt}_{\ln \dfrac{m(t_f)}{m(t_0)}} =-\dfrac{1}{c} \int_{t_0}^{t_f} \dfrac{\norm{\bm{T}(t)}_2}{m(t)}dt =-\dfrac{1}{c} \int_{0}^{1} \dfrac{\norm{\bm{T}(\tau)}_2s(\tau)}{m(\tau)}d\tau = -\dfrac{1}{c}\int_0^1\norm{\tilde{\bm{T}}(\tau)}_2 d\tau
\end{equation}

Thus, we can write the final mass as,
\begin{equation}
m(t_f) = m(t_0) \exp \left[- \dfrac{1}{c} \int_0^1 \norm{\tilde{\bm{T}}(\tau)}_2 d\tau\right].
\end{equation}

Since $c>0$, maximizing $m(t_f)$ is equivalent to minimizing $\int_0^1 \norm{\tilde{\bm{T}}(\tau)}_2 d\tau$. To our best knowledge, this is the first time this reparameterization is used for minimum-fuel trajectory optimization problems  to enable an adaptive mesh-refinement scheme.

The next step is to convexify the problem. First, an ``epigraph approach'' \cite{boyd2004convex} is used where a slack variable is introduced in place of this objective, and a second-order cone constraint is used to relax the constraint:
\begin{equation}
\begin{aligned}
    J = \int_{0}^{1} \tilde{T}(\tau) d \tau, \quad \text{with} \quad
    \norm{\bm{T}(\tau)}_2\leq \tilde{T}(\tau).
\end{aligned}
\end{equation}

The equations of motion are re-written as,
\begin{equation} \label{eq: normalized_eom}
    \bm{x}'(\tau) = s\mathbb{A}(\bm{x}) + \mathbb{B}(\bm{x}) \tilde{\bm{T}}, \quad z' = -\frac{\norm{\tilde{\bm{T}}}_2}{c}
\end{equation}

For $z'$ in Eq. \eqref{eq: normalized_eom}, we also use the slack variable, changing it to $z' = -\tilde{T}/c$. Now regarding the constraint on \eqref{eq:ocp_convex_c}, with the new $\tilde{T}$ acceleration thrust we would have:
\begin{align}
    \norm{\bm{T}(t)}_2 \leq T_{\max} \underbrace{\implies}_{\bm{T}=\dfrac{\tilde{\bm{T}} e^z}{s}} \norm{\tilde{\bm{T}}}_2 \leq T_{\max}se^{-z}
\end{align}

We can then use linearization around references $\hat{z} , \hat{s}$ to linearize this constraint into:
\begin{align}    
    T_{\max}se^{-z} &\approx \cancel{T_{\max}\hat{s}e^{-\hat{z}}} + (s - \cancel{\hat{s}})T_{\max} e^{-\hat{z}} - (z - \hat{z}) T_{\max}\hat{s}e^{-\hat{z}}= T_{\max}e^{-\hat{z}}(s - \hat{s}(z- \hat{z}))
\end{align}

Next, we need to linearize the dynamics $\bm{f}(\bm{x},\tilde{\bm{T}},s)$ around references $\hat{\bm{x}},\hat{\tilde{\bm{T}}},\hat{s}$. We follow \eqref{eq:linear_approx}, except that now we also have to take $s$ into account:
\begin{subequations}
\begin{align}
&\dot{\bm{x}}(\tau) = \bm{f}(\bm{x}(\tau), \tilde{\bm{T}}(\tau),s(\tau)) \approx A_L(\tau) \bm{x}(\tau) + B_L(\tau) \tilde{\bm{T}}(\tau) + \bm{d}_L(\tau)s(\tau)+ \bm{e}(\tau), \\
&A_L(\tau) \coloneq \left. \dfrac{\partial \bm{f}(\bm{x}, \tilde{\bm{T}},s)}{\partial \bm{x}} \right|_{\hat{\bm{x}}(\tau), \hat{\tilde{\bm{T}}}(\tau),\hat{s}(\tau)},~B_L(\tau) \coloneq \left. \dfrac{\partial \bm{f}(\bm{x}, \tilde{\bm{T}},s)}{\partial \tilde{\bm{T}}} \right|_{\hat{\bm{x}}(\tau), \hat{\tilde{\bm{T}}}(\tau),\hat{s}(\tau)},\\
&\bm{d}_L(\tau) \coloneq \left. \dfrac{\partial \bm{f}(\bm{x}, \bm{\tilde{\bm{T}}},s)}{\partial s} \right|_{\hat{\bm{x}}(\tau), \hat{\tilde{\bm{T}}}(\tau)}\\
&\bm{e}(\tau) \coloneq \bm{f}(\hat{\bm{x}}(\tau), \hat{\tilde{\bm{T}}}(\tau),\hat{s}(\tau)) - A_L(\tau) \hat{\bm{x}}(\tau) - B_L(\tau) \hat{\tilde{\bm{T}}}(\tau) - \bm{d}_L(\tau) \hat{s}(\tau).
\end{align}
\label{eq:linear_approx_2}
\end{subequations}

By using STM we will get:
\begin{subequations}
\begin{align}
&\bm{x}_{k+1} = \bar{A}_k \bm{x}_k + \bar{B}_k \bm{u}_k + \bar{C}_k \tilde{\bm{T}}_{k+1} + \bar{\bm{d}}_k s_k + \bar{\bm{e}}_k, \quad \forall k \in \bar{\mathcal{K}}, \\
&\bar{A}_k \coloneq \Phi_A(\tau_{k+1}, \tau_k),\\
&\bar{B}_k \coloneq \int_{\tau_k}^{\tau_{k+1}} \Phi_A(\tau_{k+1}, \tau) B_L(\tau) \alpha_k(\tau) d\tau, \label{eq:conv_action_2} \bar{C}_k \coloneq \int_{\tau_k}^{\tau_{k+1}} \Phi_A(\tau_{k+1} , \tau) B_L(\tau) \beta_k(\tau) d\tau,\\
&\bar{\bm{d}}_k \coloneq \int_{\tau_k}^{\tau_{k+1}} \Phi_A(\tau_{k+1}, \tau) \bm{d}_L(\tau)  d\tau, \bar{\bm{e}}_k \coloneq \int_{\tau_k}^{\tau_{k+1}} \Phi_A(\tau_{k+1}, \tau) \bm{e}(\tau) d\tau,
\end{align}
\label{eq:discrete_dynamics_2}
\end{subequations}

One practical concern is the issue of \textit{artificial infeasibility} \cite{malyuta2021convex,liu2017survey,mao2018successive,nurre2022comparison}. The artificial infeasibility occurs when, despite the existence of a feasible solution to the original nonlinear dynamics, the problem becomes infeasible after linearization is applied. To address this, slack variables are introduced to help the convex optimization process find a solution. Essentially, the slack variables play the role of ``virtual'' controls and prevent the potential infeasibility caused by the linearization. 

The introduced slack variables are then penalized (with the use of a properly chosen norm and penalty factor) to ensure that, when a feasible trajectory exists, the slack variables approach zero. Eventually and on a converged solution, the penalty term will be effectively negligible, meaning that the obtained states and controls are accurate enough to satisfy the equations of motion, and therefore the solution is a dynamically feasible one.
Let  $\bm{h}_k \in \mathbb{R}^{6}\;~\forall k \in \mathcal{\bar{K}}$ denote the vector of slack variables which are utilized as,
\begin{equation}
    \bm{x}_{k+1} = \bar{A}_k \bm{x}_k + \bar{B}_k \tilde{\bm{T}}_k + \bar{C}_k \tilde{\bm{T}}_{k+1} + \bar{\bm{d}}_ks_k + \bar{\bm{e}}_k  + \bm{h}_k ,\quad \forall k \in \bar{\mathcal{K}},
\end{equation}

where the index set are defined as:
\begin{equation}
\begin{aligned}
&\bar{\mathcal{K}} \coloneq \{0,1,\dots,K-3,K-2\}.
\end{aligned}
\end{equation}

Where, K is the total number of end points that construct our segments. To penalize these slack variables, $C\sum_{k \in \bar{\mathcal{K}}} \sum_{i = 0}^5 |h_{k,i}|$ is added to the objective where $C$ is a coefficient for penalizing.

Another common issue in applying SCVX is due to linearization, in which the previously bounded dynamics (with finite thrust) may now become relatively unbounded and, in some cases, ill-defined \cite{mao2018successive}. For these reasons, it is desirable to limit the deviation from the reference state variables (within each iteration). This approach, rooted in optimization theory \cite{powell1970hybrid}, is known as the trust-region method. The idea is simply enforced by adding two new constraints as, 
\begin{align}
|\bm{x}_{k} - \hat{\bm{x}}_{k}| \leq \bm{r}^{\bm{x}}_{k},\quad 
|s_{k} - \hat{s}_{k}| \leq r^s_{k},
\end{align}
meaning that the deviation from state trajectory and time-dilation variable cannot exceed a threshold defined as $\bm{r}^{\bm{x}},r^s$ respectively. While some approaches \cite{nurre2022comparison,wang2018minimum} consider a constant threshold $\bm{r}$, in this research, we focus on an adaptive approach introduced in \cite{mao2018successive}. 

Combining all, we can write the following convex subproblem:
\begin{subequations}
\begin{align}
&\underset{\tilde{\bm{T}}_k,\bm{x}_k,z_k,\bm{h_k}}{\text{Minimize}}\; L = \dfrac{1}{K-1} \sum_{k \in \mathcal{\bar{K}}} \dfrac{\tilde{T}_k + \tilde{T}_{k+1}}{2}+ C\sum_{k \in \bar{\mathcal{K}}} \sum_{i = 0}^5 |h_{k,i}|, \label{eq:convex_cost}\\
&\text{s.t.:} \nonumber \\
&\bm{x}_{k+1} = \bar{A}_k \bm{x}_k + \bar{B}_k \tilde{\bm{T}}_k + \bar{C}_k \tilde{\bm{T}}_{k+1} + \bar{\bm{e}}_ks_k + \bar{\bm{d}}_k  + \bm{h}_k ,\quad \forall k \in \bar{\mathcal{K}},\\
&z_{k+1} = z_{k} - \dfrac{\tilde{T}_{k} + \tilde{T}_{k+1}}{2(K-1)c}, \label{eq:pmass_trapezoid}\\
&\norm{\tilde{\bm{T}}_{k'}}_2 \leq \tilde{T}_{k'}, \quad \forall k'\in \{0,1,\dots,K-2,K-1\},\\
&\tilde{T}_{k'} \leq T_{\max}e^{-\hat{z}}(s - \hat{s}(z- \hat{z})),\\
 &\bm{x}_0 = \bm{x}(t_0),\quad 
\bm{x}_{K-1}  = \bm{x}(t_f),\quad
z_0 = \ln(m_0),\\
& |\bm{x}_{k} - \hat{\bm{x}}_{k}| \leq \bm{r}^{\bm{x}}_{k},\quad |s_{k} - \hat{s}_{k}| \leq r^s_{k},\label{eq:trust_region1}\\
&s_k \geq 0, \quad \sum_{k \in \bar{\mathcal{K}}}{\dfrac{s_k}{K-1}} = t_f
\end{align}
\label{eq:convex_discretized_problem}
\end{subequations}
where linear discretization was in the objective and differential equation of the logarithm of mass. To adaptively update the trust region, we must distinguish between the actual cost, $J$, from the cost returned after optimizing Eq.~\eqref{eq:convex_cost}. By ``actual cost,'' we mean the result when the obtained thrust profile is applied to each segment of the trajectory (divided by discrete starting points throughout the entire trajectory). While the thrust cost remains identical, the discrepancy between the achieved states and the desired ones differs from the slack variables (which are introduced to track this difference in the linearized version). Thus, for each segment, we can define:
\begin{equation}
    \bm{E}_k = \left| \underbrace{\bm{x}_{k} + \int_{\tau_k}^{\tau_{k+1}} \bm{f}(\bm{x}(\tau),\bm{\tilde{\bm{T}}}(\tau),s(\tau)) \, d\tau}_{\text{solution of the nonlinear dynamics}} \quad - \underbrace{\bm{x}_{k+1}}_{\text{solution of the convex optimization}}\right|, \quad \forall k \in \bar{\mathcal{K}},
\end{equation}
\noindent where $\bm{E}_k$ represents the discrepancy between the state produced by convex optimization and the actual state obtained by propagating the nonlinear dynamics, $\bm{f}$, from the point $\bm{x}_k$. Consequently, the nonlinear cost is defined as:
\begin{equation}
    J = \int_{0}^{1}\tilde{T}(\tau) \, d \tau + C\sum_{k \in \bar{\mathcal{K}}}\sum_{i=0}^{5} E_{k,i}. \label{eq:actual_nonlinear_cost}
\end{equation}

By monitoring the linear cost $L$ (given in Eq.~\eqref{eq:convex_cost}), and the nonlinear cost, $J$ (given in Eq.~\eqref{eq:actual_nonlinear_cost}), the adaptive-mesh SCVX algorithm is defined in Algorithm \ref{algo:scvx}. After each iteration of optimization, the algorithm monitors the predicted improvement (i.e., the difference between the previous nonlinear cost and the obtained convex cost) and the actual nonlinear improvement. Note that $\bm{r}$ is concatenation of $\bm{r}^{\bm{x}},r^s$.

\begin{algorithm}
\caption{The adaptive-mesh SCVX Algorithm}
\begin{algorithmic}[1]
\Procedure{SCVX}{$\hat{\bm{x}},\hat{\tilde{\bm{T}}}, \hat{\bm{z}},\hat{\bm{s}},\epsilon_{\text{tol}}$}
\State \textbf{input} Select initial state $\hat{\bm{x}} \in \mathbb{R}^{(K-1)\times 6}$, control $\hat{\tilde{\bm{T}}} \in \mathbb{R}^{(K-1) \times 3}$, logarithm of mass $\hat{\bm{z}} \in \mathbb{R}^{N-1}$, and $\hat{\bm{s}} \in \mathbb{R}^{K-1}$. Initialize trust region vector $\bm{r^x} , r^s > 0$. Select parameters $0 < \rho_0 < \rho_1 < \rho_2 < 1$ and $\alpha > 1$, $\beta > 1$.
\While{not converged, i.e., $\Delta L > \epsilon_{\text{tol}}$}
    \State \textbf{step 1} Solve Problem ~\eqref{eq:convex_discretized_problem} at $(\hat{\bm{x}}, \hat{\tilde{\bm{T}}}, \hat{\bm{z}},\hat{\bm{s}})$ to get an optimal solution $(\bm{x}, \tilde{\bm{T}}, \bm{z},\bm{s})$.
    \State \textbf{step 2} Compute the actual change in the penalty cost Eq.~\eqref{eq:actual_nonlinear_cost}:
    \begin{equation}
        \Delta J = J(\hat{\bm{x}}, \hat{\tilde{\bm{T}}}, \hat{\bm{z}},\hat{\bm{s}}) - J(\bm{x}, \tilde{\bm{T}}, \bm{z},\bm{s}),
    \end{equation}
    and the predicted change by the convex cost ~\eqref{eq:convex_cost}:
    \begin{equation}
        \Delta L = J(\hat{\bm{x}}, \hat{\tilde{\bm{T}}}, \hat{\bm{z}},\hat{\bm{s}}) - L(\bm{x}, \tilde{\bm{T}}, \bm{z},\bm{s}).
    \end{equation}
        \State compute the ratio
        \begin{equation}
            \rho := \frac{\Delta J}{\Delta L}.
        \end{equation}
    \State \textbf{step 3}
    \If{$\rho < \rho_0$}
        \State reject this step, contract the trust region radius, i.e., $\bm{r} \gets \bm{r} / \alpha$ and go back to step 1;
    \Else
        \State accept this step, i.e., $\hat{\bm{x}} \gets \bm{x}$, $\hat{\tilde{\bm{T}}} \gets \bm{\tilde{T}}$, $\hat{\bm{z}} \gets \bm{z}$,$\hat{\bm{s}} \gets \bm{s}$, and update the trust region radius $\bm{r}^{k+1}$ by
        \begin{equation}
            \bm{r} \gets \begin{cases}
                \bm{r}/\alpha, & \text{if } \rho < \rho_1; \\
                \bm{r}, & \text{if } \rho_1 \leq \rho < \rho_2; \\
                \beta \bm{r}, & \text{if } \rho_2 \leq \rho.
            \end{cases}
        \end{equation}
    \EndIf
    \State go back to step 1.
\EndWhile
\State \textbf{return} $(\hat{\bm{x}}, \hat{\tilde{\bm{T}}}, \hat{\bm{z}},\hat{\bm{s}})$.
\EndProcedure
\end{algorithmic}
\label{algo:scvx}
\end{algorithm}

\section{Enhancing Trust Region Using Nonlinearity Index}

\begin{figure} [hb!]
    \centering
    \includegraphics[width=0.8\linewidth]{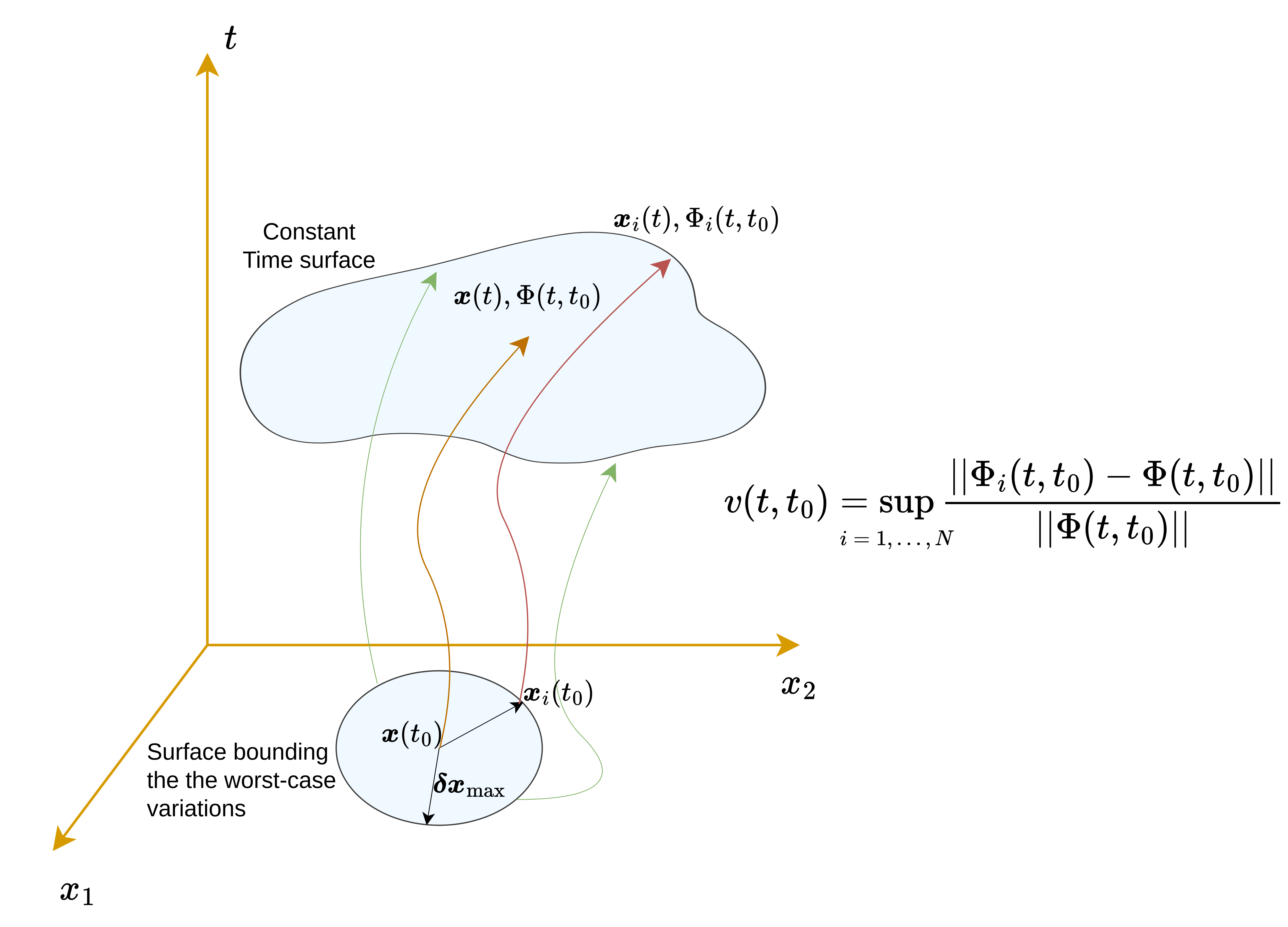}
    \caption{An illustration showcasing nonlinearity index for a two-dimensional problem.}
    \label{fig:nonlinearity_idx_demo}
\end{figure}

In \cite{bernardini2024state}, it is proposed that, since inaccuracies from linearization clearly arise from nonlinearities, measuring these nonlinearities could potentially allow for adjusting the trust region accordingly—making it more conservative (by tightening the trust-region radius) in highly nonlinear areas and more generous (by relaxing the the trust-region radius) in more linear regions. Therefore, the study suggests using the successful nonlinearity index introduced by Junkins in \cite{junkins1997karman,junkins2004nonlinear}. The STM, $\Phi_A(\tau,\tau_0)$, contains the information of zero-input evolution of trajectory, relating the initial state to a final state at time $t$. Clearly, if the dynamics are linear, the STM will be constant given any initial state. On the other hand, the higher the deviation, the higher the nonlinearity. 
Given this intuition, one can write the nonlinearity index as,
\begin{equation}
v(\tau,\tau_0) \coloneq 
\sup_{i = 1,\dots,N}\dfrac{\norm{\Phi_i(\tau,\tau_0) - \Phi(\tau,\tau_0)}}{\norm{\Phi(\tau,\tau_0)}},
\label{eq:nonlinearity_index}
\end{equation}
where given an initial state, $x_0$, and using Eq.~\eqref{eq:stm_ode_cvx}, the STM $\Phi(\tau,\tau_0)$ is propagated until time $\tau$. As for $\Phi_i$, the initial state is perturbed (see Fig. \ref{fig:nonlinearity_idx_demo}) over $N$ sample perturbations as,
\begin{equation}
\bm{x}_i(\tau_0) = \bm{x}(\tau_0) + \delta \bm{x}_i(\tau_0), \quad \text{with} \quad \norm{\delta \bm{x}_i(\tau_0)} = \delta x_{\max}, \text{for}~ i=1,\cdots,N.
\label{eq:perturbed_init}
\end{equation}


The same set of equations is then propagated forward in time. The two resulting matrices are subtracted, and the norm of this difference serves as the nonlinearity index, after being normalized by the magnitude of $\Phi$ to eliminate the influence of the first-order linear transformation. Finally, we sample from these perturbations and select the maximum of the collected nonlinearity indices. Although sampling can be used to compute the nonlinearity matrix, \cite{fossa2022multifidelity} suggests that this issue can be avoided by approximating the STM using State Transition Tensor (STT) Models \cite{younes2012high, bani2016high, bani2019exact}. While the STM provides a first-order approximation of zero-input state evolution, higher-order state transition tensors can also be employed. Define:
\begin{equation}
\Lambda_{IJK} (\tau,\tau_0) \coloneq \dfrac{\partial^2 \bm{x}_I(\tau)}{\partial \bm{x}_J (\tau_0)\partial \bm{x}_K(\tau_0)} = \dfrac{\partial \Phi_{IJ}(\tau,\tau_0)}{\partial \bm{x}_K (\tau_0)},
\label{eq:def_state_transition_tensor}
\end{equation} 
where $\Lambda_{IJK} \in \mathbb{R}^{IJK}$ denotes the second-order STT. The method used to derive the differential equations for the STM can similarly be applied here to compute the state transition tensor. The only distinction is using tensor derivative conventions introduced in \cite{tafazzol2024enhanced}.

\begin{subequations}
\begin{align}
\intertext{Starting from the original ordinary differential equations (ODEs):}
\bm{x}_{I}(\tau_1) &= \bm{x}_{I}(\tau_0) + \int_{\tau_0}^{\tau_1} \bm{f}_{I}(\bm{x}(\tau), \bm{u}(\tau), s(\tau)) \, d\tau, \label{eq:ode_tensor_eqs_1}\\
\intertext{After taking the derivative:}
&\implies \left( \frac{\partial \bm{x}_{I}(\tau_1)}{\partial \bm{x}_{J}(\tau_0)} \right) = I_{IJ} + \int_{\tau_0}^{\tau_1} \frac{\partial \bm{f}_{I}(\bm{x}(\tau), \bm{u}(\tau), s(\tau))}{\partial \bm{x}_{D}(\tau)} \frac{\partial \bm{x}_{D}(\tau)}{\partial \bm{x}_{J}(\tau_0)} \, d\tau,\label{eq:ode_tensor_eqs_2}\\
\intertext{We form the State Transition Tensor ODE:}
&\implies \frac{\partial^{2} \bm{x}_{I}(\tau_1)}{\partial \bm{x}_{J}(\tau_0) \partial \bm{x}_{K}(\tau_0)} = \frac{\partial}{\partial \bm{x}_{K}(\tau_0)} \left( \frac{\partial \bm{x}_{I}(\tau_1)}{\partial \bm{x}_{J}(\tau_0)} \right) \label{eq:ode_tensor_eqs_3}\\
 = \int_{\tau_0}^{\tau_1} &\left[ \frac{\partial}{\partial \bm{x}_{K}(\tau_0)} \left( \frac{\partial \bm{f}_{I}(\bm{x}(\tau), \bm{u}(\tau), s(\tau))}{\partial \bm{x}_{D}(\tau)} \right) \frac{\partial \bm{x}_{D}(\tau)}{\partial \bm{x}_{J}(\tau_0)} + \frac{\partial \bm{f}_{I}(\bm{x}(\tau), \bm{u}(\tau), s(\tau))}{\partial \bm{x}_{D}(\tau)} \frac{\partial}{\partial \bm{x}_{K}(\tau_0)} \left( \frac{\partial \bm{x}_{D}(\tau)}{\partial \bm{x}_{J}(\tau_0)} \right) \right] d\tau, \label{eq:ode_tensor_eqs_4}\\
\intertext{where:}
&\frac{\partial}{\partial \bm{x}_{K}(\tau_0)} \left( \frac{\partial \bm{f}_{I}(\bm{x}(\tau), \bm{u}(\tau), s(\tau))}{\partial \bm{x}_{D}(\tau)} \right) = \frac{\partial^{2} \bm{f}_{I}(\bm{x}(\tau), \bm{u}(\tau), s(\tau))}{\partial \bm{x}_{D}(\tau) \partial \bm{x}_{E}(\tau)} \frac{\partial \bm{x}_{E}(\tau)}{\partial \bm{x}_{K}(\tau_0)}, \label{eq:ode_tensor_eqs_5}\\
&\frac{\partial}{\partial \bm{x}_{K}(\tau_0)} \left( \frac{\partial \bm{x}_{D}(\tau)}{\partial \bm{x}_{J}(\tau_0)} \right) = \frac{\partial^{2} \bm{x}_{D}(\tau)}{\partial \bm{x}_{J}(\tau_0) \partial \bm{x}_{K}(\tau_0)} \label{eq:ode_tensor_eqs_6} \implies \frac{\partial^{2} \bm{x}_{I}(\tau_1)}{\partial \bm{x}_{J}(\tau_0) \partial \bm{x}_{K}(\tau_0)} =\\
  &\int_{\tau_0}^{\tau_1} \left[ \frac{\partial^{2} \bm{f}_{I}(\bm{x}(\tau), \bm{u}(\tau), s(\tau))}{\partial \bm{x}_{D}(\tau) \partial \bm{x}_{E}(\tau)} \frac{\partial \bm{x}_{E}(\tau)}{\partial \bm{x}_{K}(\tau_0)} \frac{\partial \bm{x}_{D}(\tau)}{\partial \bm{x}_{J}(\tau_0)} + \frac{\partial \bm{f}_{I}(\bm{x}(\tau), \bm{u}(\tau), s(\tau))}{\partial \bm{x}_{D}(\tau)} \frac{\partial^{2} \bm{x}_{D}(\tau)}{\partial \bm{x}_{J}(\tau_0) \partial \bm{x}_{K}(\tau_0)} \right] d\tau, \label{eq:ode_tensor_eqs_7}\\
\intertext{Here, each term represents:}
&\Lambda_{I J K}(\tau_1) = \int_{\tau_0}^{\tau_1} \left[ H_{I D E}(\tau) \Phi_{E K}(\tau, \tau_0) \Phi_{D J}(\tau, \tau_0) + A_{I D}(\tau) \Lambda_{D J K}(\tau, \tau_0) \right] d\tau.\label{eq:ode_tensor_eqs_8}
\end{align}
\label{eq:ode_state_transition_tensor}
\end{subequations}
Starting from the system dynamics in Eq.~\eqref{eq:ode_tensor_eqs_1}, by taking the partial derivative w.r.t. $\bm{x}_J(\tau_0)$, we obtain the already mentioned ODE for the State Transition Matrix (STM), this time with specified dimensions. In Eq.~\eqref{eq:ode_tensor_eqs_2}, the resulting STM has dimensions $IJ$, hence the two factors in the integral—having dimensions $ID$ and $DJ$—should yield $IJ$ through matrix multiplication.

To derive the STT, we take another derivative in Eq.~\eqref{eq:ode_tensor_eqs_3} w.r.t. $\bm{x}_K(\tau_0)$. Note that the subscript $K$ is used only to distinguish dimensions and does not imply that $\bm{x}(\tau_0)$ is a different variable. In Eq.~\eqref{eq:ode_tensor_eqs_4}, the product rule is applied, while in Eq.~\eqref{eq:ode_tensor_eqs_5}, the chain rule is utilized. Eq.~\eqref{eq:ode_tensor_eqs_6} reverts to the definition of the second-order STT. Substituting in these results yields Eq.~\eqref{eq:ode_tensor_eqs_7}. Finally, by replacing each term with its respective representation, we arrive at Eq.~\eqref{eq:ode_tensor_eqs_8}, where $H$ is the Hessian of the dynamics (\(\dot{\bm{x}}\)) w.r.t. the state variables, $A$ is the Jacobian, $\Phi$ is the STM, and $\Lambda$ is the STT. The integral in Eq.~\eqref{eq:ode_tensor_eqs_8} contains two terms as integrands. In the first term, the factors have dimensions $IDE$, $EK$, and $DJ$, while in the second term, they have dimensions $ID$ and $DJK$. Both yield the final dimension of $IJK$. To compute these operations, the Einstein summation convention (Einsum) can be conveniently utilized using Python's NumPy:

\lstdefinestyle{python}{
    language=Python,
    basicstyle=\ttfamily\scriptsize, 
    keywordstyle=\color{blue}\bfseries,
    stringstyle=\color{red},
    commentstyle=\color{gray}\itshape,
    showstringspaces=false,
    columns=flexible,
    frame=single,
    breaklines=true,
    escapeinside={(*@}{@*)}, 
}

 \begin{lstlisting}[style=python]
Lambda_dot=
           np.einsum('IDE,EK,DJ->IJK', H, Phi, Phi) + np.einsum('ID,DJK->IJK', A, Lambda).
\end{lstlisting}

Like the STM, the STT can also be computed for each trajectory segment during the convex optimization process. In \cite{fossa2022multifidelity}, it is proposed that the perturbed STM in the numerator of Eq.~\eqref{eq:nonlinearity_index} can be approximated using the STT as follows:
\begin{equation}    
\dfrac{\norm{\Phi_i(\tau,\tau_0) - \Phi(\tau,\tau_0)}}{\norm{\Phi(\tau,\tau_0)}} \approx \dfrac{\norm{\cancel{\Phi(\tau,\tau_0)} + [\Lambda_{IJK}(\tau,\tau_0) \delta x_{i,K}]_{IJ} - \cancel{\Phi(\tau,\tau_0)}}}{\norm{\Phi(\tau,\tau_0)}},
\label{eq:approx_STM}
\end{equation}
where the tensor approximation replaces the need for directly computing the perturbed STM. The same authors in \cite{losacco2024low} assume that each entry of $\delta x$ is either $+1$ or $-1$. Under this assumption, they use the entrywise $L_1$ norm for the matrix, leading to the following expression for the nonlinearity matrix:
\begin{equation}
v(\tau,\tau_0) = \sup_{i=1,\dots,N} \dfrac{\norm{[\Lambda_{IJK}(\tau,\tau_0) \delta \bm{x}_{i,K}]_{IJ}}}{\norm{\Phi(\tau,\tau_0)}} = \dfrac{\sum_{i=1}^{n}\sum_{j=1}^{n}\sum_{k=1}^{n} |\Lambda_{ijk}(\tau,\tau_0)|}{\sum_{i=1}^{n}\sum_{j=1}^{n}|\Phi_{ij}(\tau,\tau_0)|}.
\label{eq:approx_STM2}
\end{equation}

Furthermore, since we are interested in the contribution of each state variable to the nonlinearity index, the directional nonlinearity index can be computed by considering the specific variable of interest in the last dimension as,
\begin{equation}
v_e(\tau,\tau_0) = \dfrac{\sum_{i=1}^{n}\sum_{j=1}^{n}|\Lambda_{ije}(\tau,\tau_0)|}{\sum_{i=1}^{n}\sum_{j=1}^{n}|\Phi_{ij}(\tau,\tau_0)|}.
\label{eq:directional_nonlin_index}
\end{equation}
This would be the case if the perturbation vector $\delta\bm{x}_e$ takes the form of a one hot vector: \\$\begin{bmatrix} 0 &\dots &0& \delta x & 0\dots & 0 \end{bmatrix}$. In the previous section, the trust region, $\bm{r}$ in~\eqref{eq:trust_region1} remained constant throughout the trajectory segments. Now, given this nonlinearity index for each segment and along each direction, we will arrive at $\bm{v}_{\mathcal{K}E}$, where $\mathcal{K}$ is the dimension for segments and $E$ is the dimension for each direction of the state vector. Then the trust region can be modified using the nonlinearity index accordingly as,
\begin{equation}
\tilde{\bm{r}}_{\mathcal{K}E} \gets (\dfrac{\eta}{\bm{v}})_{\mathcal{K}E} \bm{r}_{E},
\label{eq:turst_region_nonlinear}
\end{equation}
where $\eta$ is a scaling parameter, $\bm{r}$ is the trust region that was previously used with dimension $E$, and $\mathcal{K}$ is the number of discrete points. Replacing $\tilde{\bm{r}}$ in Eq.~\eqref{eq:trust_region1} with the previous trust region allows for penalizing nonlinearity while maintaining a larger trust region in more linear regions. For stability, $\gamma = \eta/\bm{v}$ is clipped between two values. Note that if the system is almost linear, this value can become enormous. In this research, we apply nonlinearity index to state variables only, but an interesting experiment in future is to assess the effect of nonlinearity index on time-dilation variable as well.

\section{Numerical Simulations}

 To assess the effectiveness of the novel proposed methods, we conduct experiments on two space trajectory benchmarks: Earth-to-Dionysus (E2D) \cite{taheri2016enhanced} and a Halo-to-Halo maneuver in the Circular Restricted Three-Body Problem (CR3BP) \cite{saloglu2024acceleration}. The experiments were implemented in Python. The CasADi library~\cite{Andersson2019} was used for system modeling and symbolic computation. Numerical integration was carried out using SciPy~\cite{2020SciPy-NMeth}. Convex subproblems were formulated in CVXPY~\cite{diamond2016cvxpy} and solved using the ECOS solver~\cite{Domahidi2013ecos}. The code is  published in \href{https://github.com/saeidtafazzol/scvx_mesh_ref_nonlinearity_idx}{\textcolor{blue}{\textit{here}}}.

 
\subsection{Earth to Dionysus}
This popular benchmark problem \cite{taheri2016enhanced} involves a spacecraft departing from Earth and performing a rendezvous with asteroid Dionysus within a fixed maneuver duration of $3534$ days. The analysis focuses solely on the heliocentric phase of flight. It is assumed that the hyperbolic excess velocities relative to both Earth (denoted with subscript '$E$') and Dionysus (denoted with subscript '$D$') are zero. The relevant parameters and boundary conditions are listed in Table~\ref{tab:scvx_parameters_earth_to_dionysus}. In this work, the Modified Equinoctial Elements (MEEs) are used for describing the spacecraft motion dynamics as described in~\cite{junkins2019exploration} and the appropriate algebraic expressions for the $\mathbb{A}$ and  $\mathbb{B}$ matrices are used.

\begin{table}[h]
    \begin{adjustwidth}{-1cm}{}
    \caption{Parameters and boundary conditions for successive convex optimization (Earth‐to‐Dionysus).}
    \scriptsize
    \begin{tabular}{ll|ll}
        \hline
        \textbf{Parameter} & \textbf{Value} & \textbf{Parameter} & \textbf{Value} \\
        \hline
        Gravitational parameter ($\mu$) & $1.32712440018\times10^{11}\,\mathrm{km}^3/\mathrm{s}^2$ 
            & Initial mass ($m_0$) & $4000\,\mathrm{kg}$ \\
        Specific impulse ($I_{\mathrm{sp}}$) & $3000\,\mathrm{s}$ 
            & Standard gravity ($g_0$) & $9.80665\,\mathrm{m/s}^2$ \\
        Astronomical unit (AU) & $149597870.69100001454 \;\mathrm{km}$ 
            & Time unit (TU) & $31536000 / (2 \pi) \;\mathrm{s}$ \\
        Maximum thrust ($T_{\max}$) & $0.32\,\mathrm{N}$ 
            &  &  \\
        State vectors &
        $\displaystyle
        \begin{aligned}
            \mathbf{r}_E &= [-3637871.081,\;147099798.784,\;-2261.441]^\top \\[-2pt]
            \mathbf{v}_E &= [-30.265097,\;-0.8486854,\;0.505\times10^{-4}]^\top \\[-2pt]
            \mathbf{r}_D &= [-302452014.884,\;316097179.632,\;82872290.0755]^\top \\[-2pt]
            \mathbf{v}_D &= [-4.533473,\;-13.110309,\;0.656163]^\top
        \end{aligned}$ 
            & Time of flight (TOF) & $3534\,\mathrm{days}$ \\
        Penalty coefficient ($C$) & $10$ 
            & Tolerance ($\epsilon_{\mathrm{tol}}$) & $1\times10^{-2}$ \\
            Trust region vector & $\begin{bmatrix}
            10, 0.1, 10, 1, 1, 10, 10
        \end{bmatrix}^{\top}$& &\\
        Step size increase factor ($\alpha$) & $1.5$ 
            & Decrease factor ($\beta$) & $1.5$ \\
        Lower threshold ($\rho_0$) & $0.2$ 
            & Middle threshold ($\rho_1$) & $0.35$ \\
        Upper threshold ($\rho_2$) & $0.8$ 
            & $\eta$ & $\dfrac{1}{30}$ \\
        Clamp range on multiplier, $\gamma$ & $[1.0,\,3.0]$ 
            &  &  \\
        \hline
    \end{tabular}
    \label{tab:scvx_parameters_earth_to_dionysus}
    \end{adjustwidth}
\end{table}

\begin{figure}[htb!]
    \centering
    \includegraphics[width=1.0\linewidth]{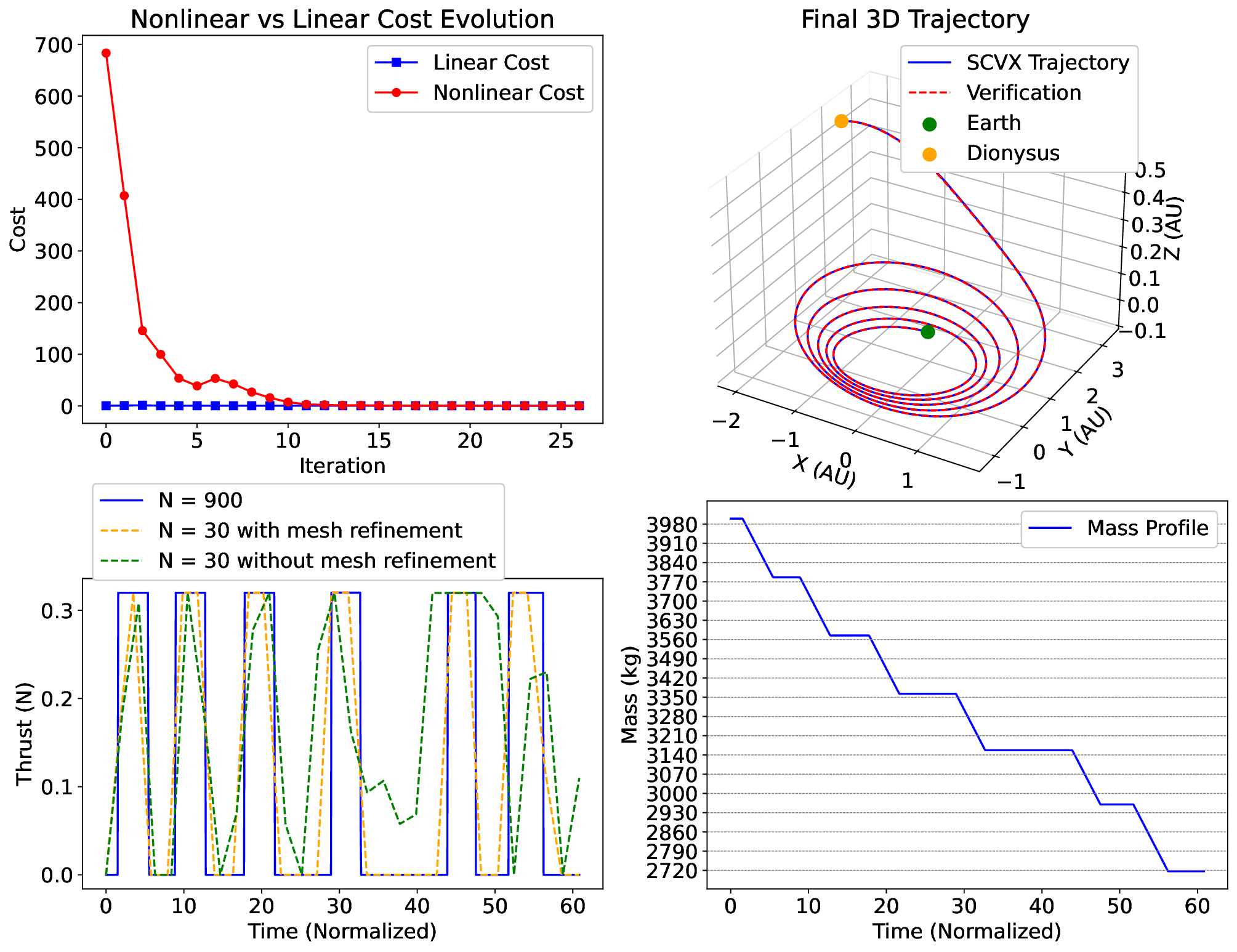}
    \caption{Summary of the Earth–Dionysus SCVX  results: (a) nonlinear vs.\ linear cost evolution over iterations; (b) optimized 3D trajectory (solid blue) with segmented‐integration verification (dashed red); (c) thrust magnitude profile over normalized time; and (d) spacecraft mass profile over normalized time.}
    \label{fig:e2d_analysis}    
\end{figure}

\begin{figure}[htb!]
    \centering
    \includegraphics[width=1.0\linewidth]{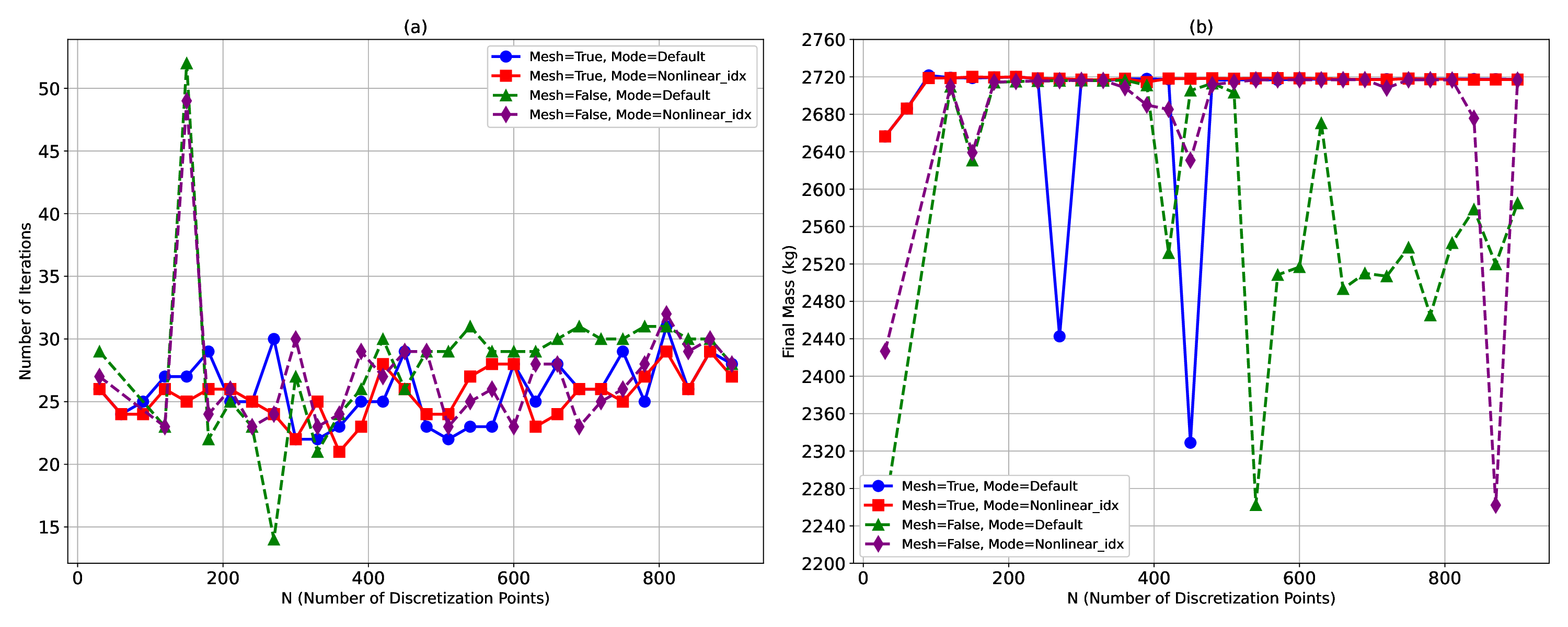}
    \caption{SCVX performance for the E2D transfer as a function of the number of discretization points $N$: (a) total iterations required for convergence for each configuration; (b) final spacecraft mass (kg) obtained after convergence.}
    \label{fig:e2d_N}    
\end{figure}

The optimization results are shown in Figure~\ref{fig:e2d_analysis}. Please note that the global minimum-fuel solution to this problem is reported in the literature by Taheri \cite{taheri2016enhanced,taheri2020many}, which consists of a maneuver in which the spacecraft makes five orbital revolutions around the Sun. To compare different approaches, we plot the convergence metrics against the number of discretization points in Figure~\ref{fig:e2d_N}. It took 38 iterations for the algorithm to converge with N = 1000 grid points. Fig.~\ref{fig:e2d_N}a shows the evolution of the linear and nonlinear costs vs. the number of iterations. Fig.~\ref{fig:e2d_N}b shows the three-dimensional trajectory associated with the solution of the convex optimization and a ``verification'' trajectory, which is obtained by numerically propagating the nonlinear dynamics using the control profile only. Fig.~\ref{fig:e2d_N}c shows the thrust magnitude profile vs. the normalized time, which consists of 12 switches. The thrust profile shows late-departure and early-arrival coast arcs \cite{taheri2020many}. Fig.~\ref{fig:e2d_N}d shows the spacecraft mass profile vs. normalized time. The final mass is 2717.117 kg, which shows very minor deviation from the known optimal solution with a final mass of 2718.37 kg, which is obtained using an indirect method. We observed that mesh refinement did not yield any improvement in the objective, and although the nonlinearity‐index heuristic reduced the number of iterations in most cases, its impact was marginal. We attribute these observations to the problem’s inherent difficulty and to the fact that, under modified equinoctial elements, the orbital dynamics exhibit significantly reduced nonlinearity \cite{junkins2004nonlinear}. Therefore, we moved on to consider a harder problem with the CR3BP dynamics.





\subsection{Halo-to-Halo in the Earth-Moon CR3BP}

\begin{figure}[htb!]
    \centering
    \includegraphics[width=1.0\linewidth]{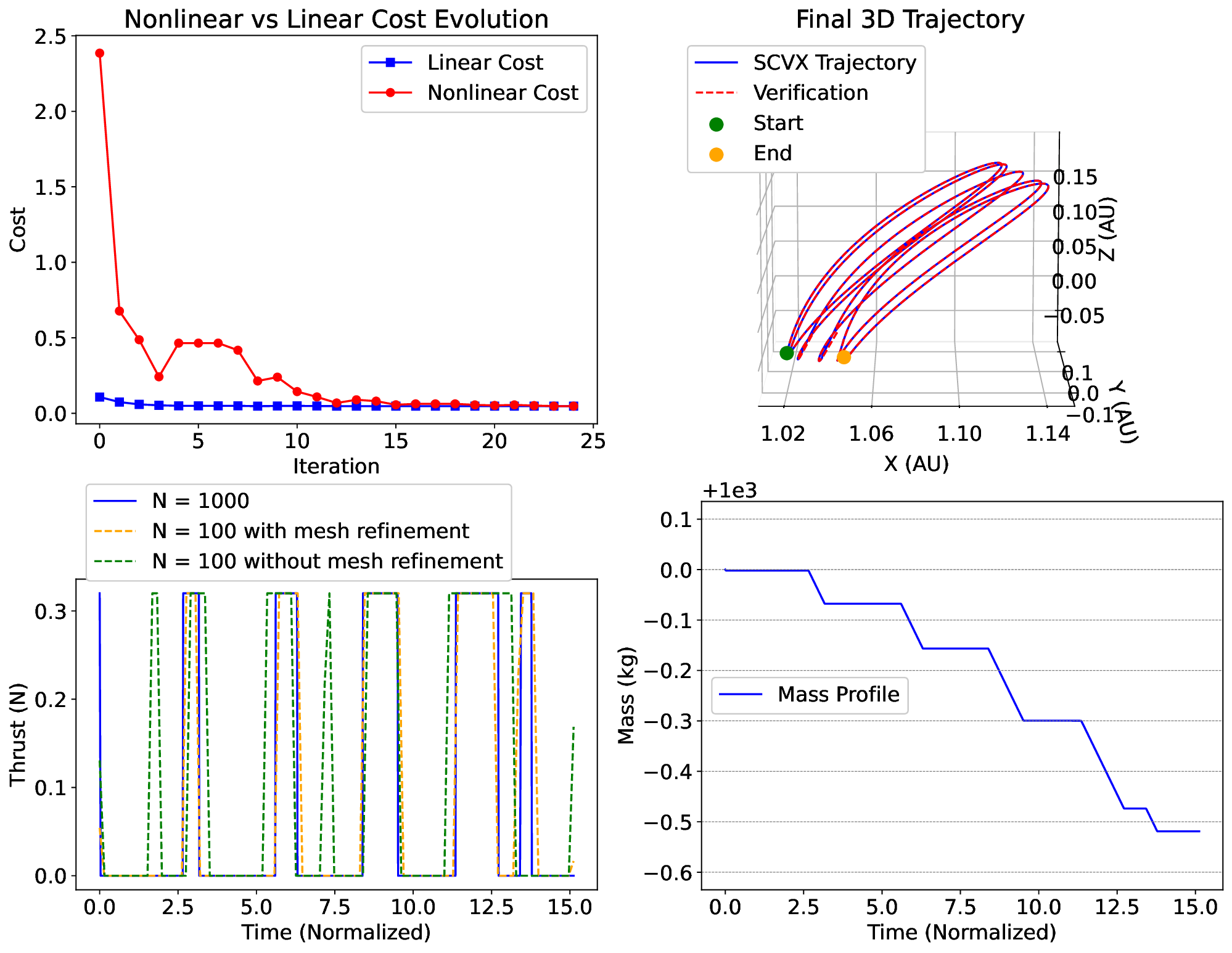}
    \caption{Four‐panel summary of the Earth–Dionysus SCVX experiment results: (a) nonlinear vs.\ linear cost evolution over iterations; (b) optimized 3D trajectory (solid blue) with segmented‐integration verification (dashed red); (c) thrust magnitude profile over normalized time; and (d) spacecraft mass profile over normalized time.}
    \label{fig:cr3bp_analysis}    
\end{figure}

The CR3BP dynamics \cite{schaub2003analytical} is a model that combines strong nonlinear gravitational dynamics with low-dimensional state space, admits well‐studied equilibrium points and periodic orbits, and captures key challenges like unstable manifolds and sensitivity to initial conditions—all within a fully normalized framework that’s ideal for testing trajectory‐optimization and feedback‐control schemes. Here, we are considering a minimum-fuel fixed-type rendezvous maneuver between two Halo orbits around the L2 points of the Earth-Moon system. The $\mathbb{A},\mathbb{B}$ for this problem are:
\begin{align}
\ell_1 &= \sqrt{(r_x+\mu)^2 + r_y^2 + r_z^2}, 
& \ell_2 &= \sqrt{(r_x+\mu-1)^2 + r_y^2 + r_z^2},\\
g_x &= r_x - (1-\mu)\frac{r_x+\mu}{\ell_1^3} - \mu\frac{r_x+\mu-1}{\ell_2^3}, 
& g_y &= r_y - (1-\mu)\frac{r_y}{\ell_1^3} - \mu\frac{r_y}{\ell_2^3},\\
g_z &= -\,(1-\mu)\frac{r_z}{\ell_1^3} - \mu\frac{r_z}{\ell_2^3}, 
& g &= \begin{bmatrix}g_x\\g_y\\g_z\end{bmatrix},\\
h_x &= 2\,v_y, 
& h_y &= -2\,v_x,\\
h_z &= 0, 
& h &= \begin{bmatrix}h_x\\h_y\\h_z\end{bmatrix},\\
\mathbb{A}^\top &= 
\begin{bmatrix}
v_x & v_y & v_z & g_x + h_x & g_y + h_y & g_z + h_z
\end{bmatrix}, \mathbb{B} = 
\begin{bmatrix}
\bm{0}_{3\times3}\\
\bm{I}_{3\times3}
\end{bmatrix}
\end{align}
The relevant parameters and boundary conditions are listed in Table~\ref{tab:scvx_parameters_cr3bp_earth_moon} and are taken from \cite{saloglu2024acceleration} except for the thrust magnitude, which is chosen to result in feasible minimum-fuel low-thrust trajectories between the chose L2 Halo orbits. The optimization results are presented in Figure~\ref{fig:cr3bp_analysis} with $N = 1000$. 

Figure~\ref{fig:cr3bp_analysis}a shows that it took 25 iterations for the algorithm to converge with N = 1000 grid points. Fig.~\ref{fig:cr3bp_analysis}a shows the evolution of the linear and nonlinear costs vs. the number of iterations. Fig.~\ref{fig:cr3bp_analysis}b shows the three-dimensional trajectory associated with the solution of the convex optimization and a ``verification'' trajectory, which is obtained by numerically propagating the nonlinear dynamics using the control profile only. Fig.~\ref{fig:cr3bp_analysis}c shows the thrust magnitude profile vs. the normalized time, which consists of 11 switches. Fig.~\ref{fig:cr3bp_analysis}d shows the spacecraft mass profile vs. normalized time. The final mass is 999.48 kg corresponding to 0.52 kg of propellant being consumed during the 15.11 days of maneuver. Please note that it is possible for other low-thrust maneuvers to exist and the reported solution may not necessarily be the globally optimal solution for the considered boundary conditions and problem parameters.  

\begin{figure}[htb!]
    \centering
    \includegraphics[width=1.0\linewidth]{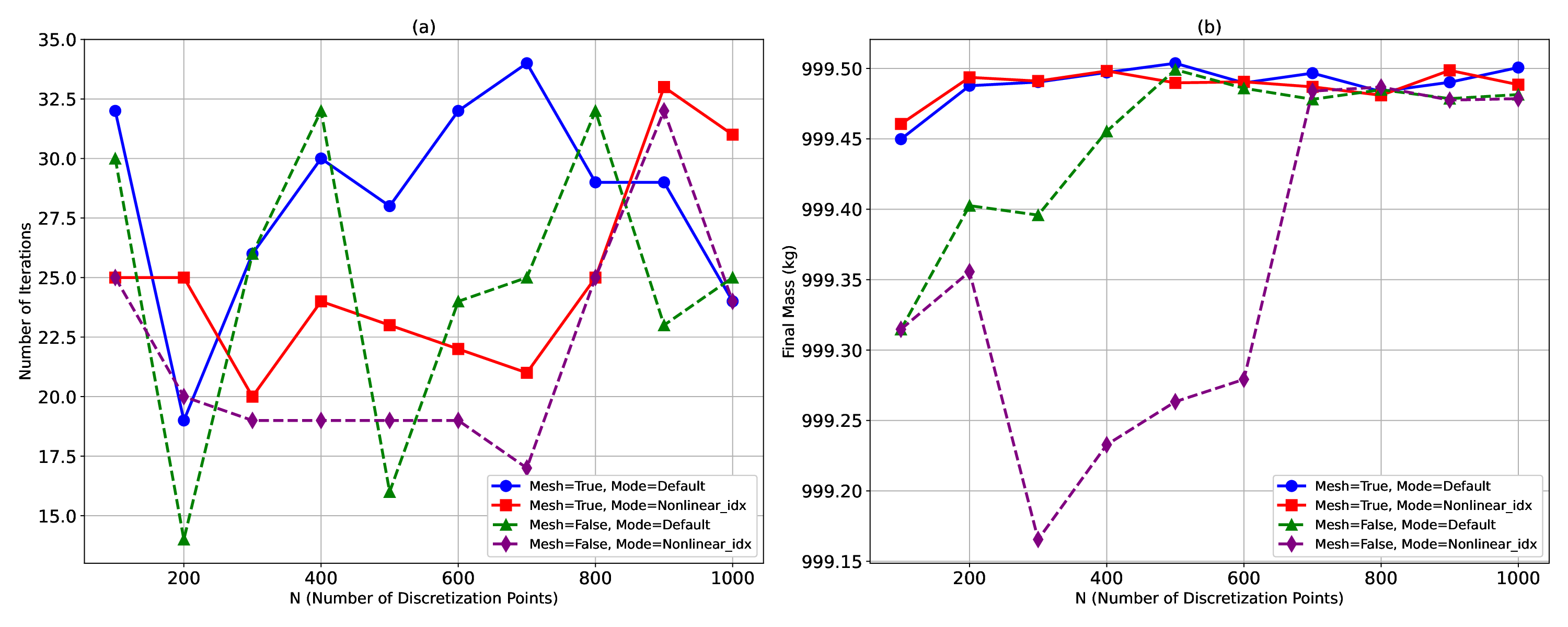}
    \vspace{-5mm}
    \caption{SCVX performance for the CR3BP transfer as a function of the number of discretization points $N$: (a) total iterations required for convergence for each configuration; (b) final spacecraft mass (kg) obtained after convergence.}
    \label{fig:cr3bp_N}    
\end{figure}

\begin{figure}[htb!]
    \centering
    \vspace{-6mm}
    \includegraphics[width=0.8\linewidth]{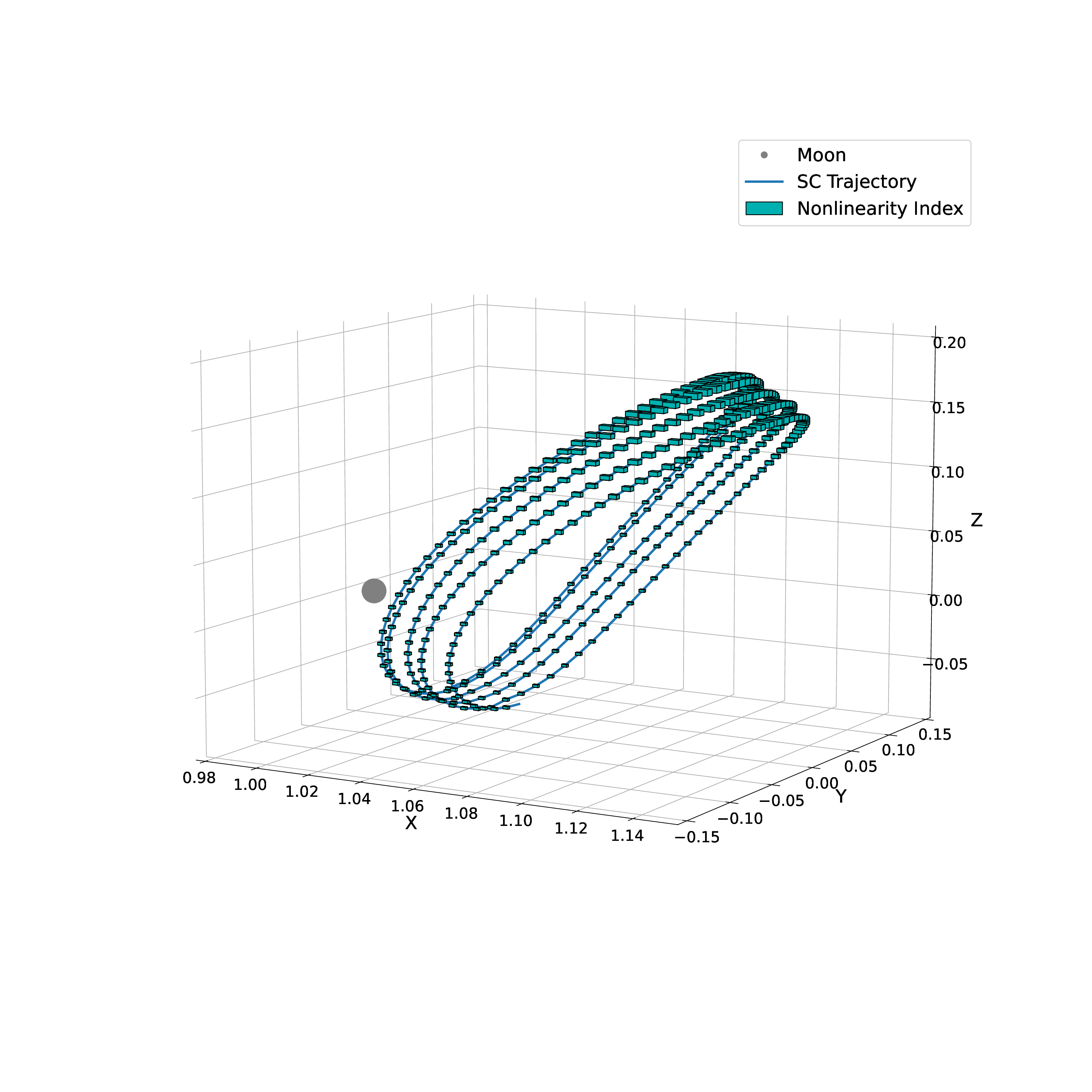}
    \vspace{-25mm}
    \caption{Halo-to-Halo maneuver demonstrating the impact of the nonlinearity index on trust region through bounding boxes distributed along the trajectory.}
    \label{fig:cr3bp_nonlinear_idx}
    \vspace{-2mm}
\end{figure}

Figure~\ref{fig:cr3bp_N} shows that, at lower discretization levels ($N$), employing mesh refinement yields a notably higher final mass compared to the uniform‐mesh case. Moreover, the iteration‐count plot on the left indicates that incorporating the nonlinearity index heuristic (purple and red curves) consistently reduces the number of iterations required for convergence. Figure \ref{fig:cr3bp_nonlinear_idx} depicts the impact of the nonlinearity index on trust region through bounding boxes distributed along a trajectory. As the spacecraft gets close to the moon, the trust region shrinks due to the increased nonlinearity, whereas the volume of the boxes increase as the spacecraft moves away from the moon that reduces the nonlinearity.  

\begin{table}[h]
    \begin{adjustwidth}{-3mm}{}
    \caption{Parameters and boundary conditions for successive convex optimization (CR3BP).}
    \scriptsize
    \begin{tabular}{ll|ll}
        \hline
        \textbf{Parameter} & \textbf{Value} & \textbf{Parameter} & \textbf{Value} \\
        \hline
        Gravitational parameter ($\mu$) & $1.21506683 \times 10^{-2}$ & Initial mass ($m_0$) & $1000 \, \text{kg}$ \\
        Specific impulse ($I_{\text{sp}}$) & $3000 \, \text{s}$ & Standard gravity ($g_0$) & $9.80665 \, \text{m/s}^2$ \\
        Length unit (LU) & $3.84405 \times 10^5 \, \text{km}$ & Time unit (TU) & $3.75676967 \times 10^5 \, \text{s}$ \\
        Maximum thrust ($T_{\text{max}}$) & $0.01$   (N) \\
        State vectors &
        $\begin{aligned}
            \mathbf{r}_i &= [1.0176, 0, -0.0699] \\
            \mathbf{v}_i &= [0, 0.4866, 0] \\
            \mathbf{r}_f &= [1.0453, 1.4989 \times 10^{-5}, -0.0755] \\
            \mathbf{v}_f &= [-5.6828 \times 10^{-5}, 0.3877, -1.0999 \times 10^{-4}]
        \end{aligned}$ & Time of flight (TOF) & $15.11$ days \\
        Penalty coefficient ($C$) & 5 \\
        Tolerance($\epsilon_{\text{tol}}$) &2.5 $\times 10^{-3}$ & $\alpha$  & $1.5$  \\
        Trust region vector & $\begin{bmatrix}
            0.1, 0.1, 0.1, 0.1, 0.1, 0.1, 0.1
        \end{bmatrix}^{\top}$
         & $\beta$ & 1.5 \\
        Lower threshold ($\rho_0$) & 0.04 & Middle threshold ($\rho_1$) & 0.2 \\
        Upper threshold ($\rho_2$) & 0.7 & $\eta$ & 0.1 \\
        Clamp range on multiplier, $\gamma$ & $[0.5, 20.0]$ \\
        \hline
    \end{tabular}
    \label{tab:scvx_parameters_cr3bp_earth_moon}
    \end{adjustwidth}
\end{table}

\section{Conclusion and Future Work}

In this paper, we have presented a novel integration of the nonlinearity index into an adaptive-mesh successive convexification (SCVX) framework for minimum-fuel, low-thrust trajectory design. Numerical results demonstrate that our approach—combining thrust–and–mass reparameterization for adaptive, time-dilation–based mesh refinement with nonlinearity-informed trust-region scaling—yields:
\begin{enumerate}
  \item \textbf{Improved convergence stability.} The nonlinearity-guided trust regions prevent overly aggressive updates in highly nonlinear segments, reducing the number of rejected SCVX iterations and accelerating overall convergence.
  \item \textbf{Enhanced solution quality at coarse discretizations.} In the CR3BP benchmark, mesh refinement guided by the nonlinearity index yielded higher final spacecraft mass at low node counts, demonstrating that strategically reallocating nodes to nonlinear regions preserves fidelity with fewer decision points.
\end{enumerate}

Future work will explore several directions. First, applying the nonlinearity index to the time-dilation variable could further optimize node placement in coasting–thrust switching regions. Second, incorporating higher-order state-transition tensors for real-time nonlinearity estimation promises even tighter trust-region control without sampling perturbations. Finally, extending this scheme to more difficult space missions with multiple gravity-assist maneuvers \cite{nurre2021application} will test its versatility.

\bibliographystyle{AAS_publication}   
\bibliography{references}   

@article{taheri2020many,
  title={How many impulses redux},
  author={Taheri, Ehsan and Junkins, John L},
  journal={The Journal of the Astronautical Sciences},
  volume={67},
  number={2},
  pages={257--334},
  year={2020},
  publisher={Springer}
}

@article{mao2018successive,
  title={Successive convexification: A superlinearly convergent algorithm for non-convex optimal control problems},
  author={Mao, Yuanqi and Szmuk, Michael and Xu, Xiangru and A{\c{c}}ikmese, Beh{\c{c}}et},
  journal={arXiv preprint arXiv:1804.06539},
  year={2018}
}

@inproceedings{nurre2022comparison,
  title={Comparison of indirect and convex-based methods for low-thrust minimum-fuel trajectory optimization},
  author={Nurre, Nicholas P and Taheri, Ehsan},
  booktitle={2022 AAS/AIAA Astrodynamics Specialist Conference, AAS 22},
  volume={782},
  pages={21},
  year={2022}
}

@article{conway2010spacecraft,
  title={Spacecraft trajectory optimization using direct transcription and nonlinear programming},
  author={Conway, Bruce A and Paris, Stephen W},
  journal={Spacecraft trajectory optimization},
  volume={29},
  pages={37},
  year={2010},
  publisher={Cambridge University Press Cambridge, UK}
}

@article{wang2024survey,
  title={A survey on convex optimization for guidance and control of vehicular systems},
  author={Wang, Zhenbo},
  journal={Annual Reviews in Control},
  volume={57},
  pages={100957},
  year={2024},
  publisher={Elsevier}
}

@article{liu2017survey,
  title={Survey of convex optimization for aerospace applications},
  author={Liu, Xinfu and Lu, Ping and Pan, Binfeng},
  journal={Astrodynamics},
  volume={1},
  pages={23--40},
  year={2017},
  publisher={Springer}
}

@article{taheri2016enhanced,
  title={Enhanced smoothing technique for indirect optimization of minimum-fuel low-thrust trajectories},
  author={Taheri, Ehsan and Kolmanovsky, Ilya and Atkins, Ella},
  journal={Journal of Guidance, Control, and Dynamics},
  volume={39},
  number={11},
  pages={2500--2511},
  year={2016},
  publisher={American Institute of Aeronautics and Astronautics}
}

@article{taheri2018generic,
  title={Generic smoothing for optimal bang-off-bang spacecraft maneuvers},
  author={Taheri, Ehsan and Junkins, John L},
  journal={Journal of Guidance, Control, and Dynamics},
  volume={41},
  number={11},
  pages={2470--2475},
  year={2018},
  publisher={American Institute of Aeronautics and Astronautics}
}

@article{arya2021electric,
  title={Electric thruster mode-pruning strategies for trajectory-propulsion co-optimization},
  author={Arya, Vishala and Taheri, Ehsan and Junkins, John},
  journal={Aerospace Science and Technology},
  volume={116},
  pages={106828},
  year={2021},
  publisher={Elsevier}
}

@article{kovryzhenko2023vectorized,
  title={Vectorized Trigonometric Regularization for Singular Control Problems with Multiple State Path Constraints},
  author={Kovryzhenko, Yevhenii and Taheri, Ehsan},
  journal={The Journal of the Astronautical Sciences},
  volume={71},
  number={1},
  pages={1},
  year={2023},
  publisher={Springer}
}

@article{mall2020uniform,
  title={Uniform trigonometrization method for optimal control problems with control and state constraints},
  author={Mall, Kshitij and Grant, Michael J and Taheri, Ehsan},
  journal={Journal of Spacecraft and Rockets},
  volume={57},
  number={5},
  pages={995--1007},
  year={2020},
  publisher={American Institute of Aeronautics and Astronautics}
}

@article{trelat2012optimal,
  title={Optimal control and applications to aerospace: some results and challenges},
  author={Tr{\'e}lat, Emmanuel},
  journal={Journal of Optimization Theory and Applications},
  volume={154},
  pages={713--758},
  year={2012},
  publisher={Springer}
}

@article{aziz2018low,
  title={Low-thrust many-revolution trajectory optimization via differential dynamic programming and a sundman transformation},
  author={Aziz, Jonathan D and Parker, Jeffrey S and Scheeres, Daniel J and Englander, Jacob A},
  journal={The Journal of the Astronautical Sciences},
  volume={65},
  pages={205--228},
  year={2018},
  publisher={Springer}
}

@book{schaub2003analytical,
  title={Analytical mechanics of space systems},
  author={Schaub, Hanspeter and Junkins, John L},
  year={2003},
  publisher={Aiaa}
}

@inproceedings{nurre2021application,
  title={Application of Finite Fourier Series for Spacecraft Trajectory Design with Multiple Gravity-Assist Maneuvers},
  author={Nurre, Nicholas Paul and Taheri, Ehsan},
  booktitle={2021 Spring Southeastern Virtual Sectional Meeting},
  organization={AMS}
}

@article{sowell2024eclipse,
  title={Eclipse-conscious low-thrust trajectory optimization using pseudospectral methods and control smoothing techniques},
  author={Sowell, Samuel and Taheri, Ehsan},
  journal={Journal of Spacecraft and Rockets},
  volume={61},
  number={3},
  pages={900--907},
  year={2024},
  publisher={American Institute of Aeronautics and Astronautics}
}

@article{saloglu2024acceleration,
  title={Acceleration-Based Switching Surfaces for Impulsive Trajectory Design Between Cislunar Libration Point Orbits},
  author={Saloglu, Keziban and Taheri, Ehsan},
  journal={The Journal of the Astronautical Sciences},
  volume={71},
  number={2},
  pages={13},
  year={2024},
  publisher={Springer}
}

@article{junkins2004nonlinear,
  title={How nonlinear is it? A tutorial on nonlinearity of orbit and attitude dynamics},
  author={Junkins, John L and Singla, Puneet},
  journal={The Journal of the Astronautical Sciences},
  volume={52},
  number={1},
  pages={7--60},
  year={2004},
  publisher={Springer}
}

@article{koeppen2019fast,
  title={Fast mesh refinement in pseudospectral optimal control},
  author={Koeppen, N and Ross, I Michael and Wilcox, Lucas C and Proulx, Ronald J},
  journal={Journal of Guidance, Control, and Dynamics},
  volume={42},
  number={4},
  pages={711--722},
  year={2019},
  publisher={American Institute of Aeronautics and Astronautics}
}

@article{betts1998mesh,
  title={Mesh refinement in direct transcription methods for optimal control},
  author={Betts, John T and Huffman, William P},
  journal={Optimal Control Applications and Methods},
  volume={19},
  number={1},
  pages={1--21},
  year={1998},
  publisher={Wiley Online Library}
}

@article{wang2018minimum,
  title={Minimum-fuel low-thrust transfers for spacecraft: A convex approach},
  author={Wang, Zhenbo and Grant, Michael J},
  journal={IEEE Transactions on Aerospace and Electronic Systems},
  volume={54},
  number={5},
  pages={2274--2290},
  year={2018},
  publisher={IEEE}
}

@inproceedings{taheri2021optimization,
  title={Optimization of many-revolution minimum-time low-thrust trajectories using sundman transformation},
  author={Taheri, Ehsan},
  booktitle={AIAA Scitech 2021 Forum},
  pages={1343},
  year={2021}
}

@article{lev2019technological,
  title={The technological and commercial expansion of electric propulsion},
  author={Lev, Dan and Myers, Roger M and Lemmer, Kristina M and Kolbeck, Jonathan and Koizumi, Hiroyuki and Polzin, Kurt},
  journal={Acta Astronautica},
  volume={159},
  pages={213--227},
  year={2019},
  publisher={Elsevier}
}

@inproceedings{whiffen2006mystic,
  title={Mystic: Implementation of the static dynamic optimal control algorithm for high-fidelity, low-thrust trajectory design},
  author={Whiffen, Gregory},
  booktitle={AIAA/AAS Astrodynamics Specialist Conference and Exhibit},
  pages={6741},
  year={2006}
}

@article{ellison2018application,
  title={Application and analysis of bounded-impulse trajectory models with analytic gradients},
  author={Ellison, Donald H and Conway, Bruce A and Englander, Jacob A and Ozimek, Martin T},
  journal={Journal of Guidance, Control, and Dynamics},
  volume={41},
  number={8},
  pages={1700--1714},
  year={2018},
  publisher={American Institute of Aeronautics and Astronautics}
}

@article{betts1998survey,
  title={Survey of numerical methods for trajectory optimization},
  author={Betts, John T},
  journal={Journal of guidance, control, and dynamics},
  volume={21},
  number={2},
  pages={193--207},
  year={1998}
}

@article{junkins2019exploration,
  title={Exploration of alternative state vector choices for low-thrust trajectory optimization},
  author={Junkins, John L and Taheri, Ehsan},
  journal={Journal of Guidance, Control, and Dynamics},
  volume={42},
  number={1},
  pages={47--64},
  year={2019},
  publisher={American Institute of Aeronautics and Astronautics}
}

@article{taheri2023l2,
  title={L2 Norm-Based Control Regularization for Solving Optimal Control Problems},
  author={Taheri, Ehsan and Li, Nan},
  journal={IEEE Access},
  volume={11},
  pages={125959--125971},
  year={2023},
  publisher={IEEE}
}

@article{petukhov2019application,
  title={Application of the angular independent variable and its regularizing transformation in the problems of optimizing low-thrust trajectories},
  author={Petukhov, VG},
  journal={Cosmic research},
  volume={57},
  number={5},
  pages={351--363},
  year={2019},
  publisher={Springer}
}

@article{meng2019low,
  title={Low-thrust minimum-fuel trajectory optimization using multiple shooting augmented by analytical derivatives},
  author={Meng, Yazhe and Zhang, Hao and Gao, Yang},
  journal={Journal of Guidance, Control, and Dynamics},
  volume={42},
  number={3},
  pages={662--677},
  year={2019},
  publisher={American Institute of Aeronautics and Astronautics}
}

@ARTICLE{2020SciPy-NMeth,
  author  = {Virtanen, Pauli and Gommers, Ralf and Oliphant, Travis E. and
            Haberland, Matt and Reddy, Tyler and Cournapeau, David and
            Burovski, Evgeni and Peterson, Pearu and Weckesser, Warren and
            Bright, Jonathan and {van der Walt}, St{\'e}fan J. and
            Brett, Matthew and Wilson, Joshua and Millman, K. Jarrod and
            Mayorov, Nikolay and Nelson, Andrew R. J. and Jones, Eric and
            Kern, Robert and Larson, Eric and Carey, C J and
            Polat, {\.I}lhan and Feng, Yu and Moore, Eric W. and
            {VanderPlas}, Jake and Laxalde, Denis and Perktold, Josef and
            Cimrman, Robert and Henriksen, Ian and Quintero, E. A. and
            Harris, Charles R. and Archibald, Anne M. and
            Ribeiro, Ant{\^o}nio H. and Pedregosa, Fabian and
            {van Mulbregt}, Paul and {SciPy 1.0 Contributors}},
  title   = {{{SciPy} 1.0: Fundamental Algorithms for Scientific
            Computing in Python}},
  journal = {Nature Methods},
  year    = {2020},
  volume  = {17},
  pages   = {261--272},
  adsurl  = {https://rdcu.be/b08Wh},
  doi     = {10.1038/s41592-019-0686-2},
}

@article{powell1970hybrid,
  title={A hybrid method for nonlinear equations},
  author={Powell, Michael JD},
  journal={Numerical methods for nonlinear algebraic equations},
  pages={87--161},
  year={1970},
  publisher={Gordon and Breach}
}

@book{boyd2004convex,
  title={Convex optimization},
  author={Boyd, Stephen and Vandenberghe, Lieven},
  year={2004},
  publisher={Cambridge university press}
}

@article{acikmese2007convex,
  title={Convex programming approach to powered descent guidance for mars landing},
  author={Acikmese, Behcet and Ploen, Scott R},
  journal={Journal of Guidance, Control, and Dynamics},
  volume={30},
  number={5},
  pages={1353--1366},
  year={2007}
}

@inproceedings{szmuk2018successive,
  title={Successive convexification for 6-dof mars rocket powered landing with free-final-time},
  author={Szmuk, Michael and Acikmese, Behcet},
  booktitle={2018 AIAA Guidance, Navigation, and Control Conference},
  pages={0617},
  year={2018}
}

@article{malyuta2021convex,
  title={Convex optimization for trajectory generation},
  author={Malyuta, Danylo and Reynolds, Taylor P and Szmuk, Michael and Lew, Thomas and Bonalli, Riccardo and Pavone, Marco and Acikmese, Behcet},
  journal={arXiv preprint arXiv:2106.09125},
  year={2021}
}

@article{diamond2016cvxpy,
  author  = {Steven Diamond and Stephen Boyd},
  title   = {{CVXPY}: {A} {P}ython-embedded modeling language for convex optimization},
  journal = {Journal of Machine Learning Research},
  year    = {2016},
  volume  = {17},
  number  = {83},
  pages   = {1--5},
}

@article{bernardini2024state,
  title={State-dependent trust region for successive convex programming for autonomous spacecraft},
  author={Bernardini, Nicol{\`o} and Baresi, Nicola and Armellin, Roberto},
  journal={Astrodynamics},
  pages={1--23},
  year={2024},
  publisher={Springer}
}

@article{junkins1997karman,
  title={von karman lecture: Adventures on the interface of dynamics and control},
  author={Junkins, John L},
  journal={Journal of Guidance, Control, and Dynamics},
  volume={20},
  number={6},
  pages={1058--1071},
  year={1997}
}

@inproceedings{fossa2022multifidelity,
  title={Multifidelity orbit uncertainty propagation using Taylor polynomials},
  author={Foss{\`a}, Alberto and Armellin, Roberto and Delande, Emmanuel and Losacco, Matteo and Sanfedino, Francesco},
  booktitle={AIAA SciTech 2022 Forum},
  pages={0859},
  year={2022}
}

@inproceedings{younes2012high,
  title={High-order uncertainty propagation enabled by computational differentiation},
  author={Younes, Ahmad Bani and Turner, James and Majji, Manoranjan and Junkins, John},
  booktitle={Recent Advances in Algorithmic Differentiation},
  pages={251--260},
  year={2012},
  organization={Springer}
}

@inproceedings{bani2016high,
  title={High order state transition tensors of perturbed orbital motion using computational differentiation},
  author={Bani Younes, A and Turner, J},
  booktitle={Proceedings of the 26th AAS/AIAA Space Flight Mechanics Meeting, Napa, CA, USA, AAS},
  pages={16--342},
  year={2016}
}

@article{bani2019exact,
  title={Exact computation of high-order state transition tensors for perturbed orbital motion},
  author={Bani Younes, Ahmad},
  journal={Journal of Guidance, Control, and Dynamics},
  volume={42},
  number={6},
  pages={1365--1371},
  year={2019},
  publisher={American Institute of Aeronautics and Astronautics}
}

@article{losacco2024low,
  title={Low-Order Automatic Domain Splitting Approach for Nonlinear Uncertainty Mapping},
  author={Losacco, Matteo and Foss{\`a}, Alberto and Armellin, Roberto},
  journal={Journal of Guidance, Control, and Dynamics},
  volume={47},
  number={2},
  pages={291--310},
  year={2024},
  publisher={American Institute of Aeronautics and Astronautics}
}

@Article{Andersson2019,
  author = {Joel A E Andersson and Joris Gillis and Greg Horn
            and James B Rawlings and Moritz Diehl},
  title = {{CasADi} -- {A} software framework for nonlinear optimization
           and optimal control},
  journal = {Mathematical Programming Computation},
  volume = {11},
  number = {1},
  pages = {1--36},
  year = {2019},
  publisher = {Springer},
  doi = {10.1007/s12532-018-0139-4}
}

@INPROCEEDINGS{Domahidi2013ecos,
author={Domahidi, A. and Chu, E. and Boyd, S.},
booktitle={European Control Conference (ECC)},
title={{ECOS}: {A}n {SOCP} solver for embedded systems},
year={2013},
pages={3071-3076}
}

@article{kumagai2024adaptive,
  title={Adaptive-mesh sequential convex programming for space trajectory optimization},
  author={Kumagai, Naoya and Oguri, Kenshiro},
  journal={Journal of Guidance, Control, and Dynamics},
  volume={47},
  number={10},
  pages={2213--2220},
  year={2024},
  publisher={American Institute of Aeronautics and Astronautics}
}

@article{szmuk2020successive,
  title={Successive convexification for real-time six-degree-of-freedom powered descent guidance with state-triggered constraints},
  author={Szmuk, Michael and Reynolds, Taylor P and A{\c{c}}{\i}kme{\c{s}}e, Beh{\c{c}}et},
  journal={Journal of Guidance, Control, and Dynamics},
  volume={43},
  number={8},
  pages={1399--1413},
  year={2020},
  publisher={American Institute of Aeronautics and Astronautics}
}

@mastersthesis{tafazzol2024enhanced,
  title={Enhanced Indirect and Convex Optimization Methods for Generating Minimum-Fuel Low-Thrust Trajectories},
  author={Tafazzol, Saeid},
  year={2024}
}

\end{document}